\documentclass[twocolumn,prl,floatfix,showpacs,nofootinbib,amsmath,amssymb]{revtex4}

\usepackage{graphicx}
\usepackage{dcolumn}
\usepackage{bm}
\usepackage{color}

\def\rmB{\mathrm{B}}

\def\VEV#1{\left\langle #1 \right\rangle}

\newcommand{\bfx}{\mathbf{x}}

\newcommand{\lsim}{\mathrel{\hbox{\rlap{\lower.55ex\hbox{$\sim$}} \kern-.3em \raise.4ex \hbox{$<$}}}}
\newcommand{\gsim}{\mathrel{\hbox{\rlap{\lower.55ex\hbox{$\sim$}} \kern-.3em \raise.4ex \hbox{$>$}}}}

\def\wigner#1#2#3#4#5#6{ \left( \begin{array}{ccc} #1 & #3 & #5
\\ #2 & #4 & #6 \\ \end{array} \right)}

\begin{document}

\title{Do baryons trace dark matter in the early universe?}

\author{Daniel Grin$^1$, Olivier Dor\'e$^{2,3}$, and Marc Kamionkowski$^2$}
\affiliation{$^1$School of Natural Sciences, Institute for Advanced
     Study, Princeton, NJ 08540}
\affiliation{$^2$California Institute of Technology, Mail Code 350-17,
     Pasadena, CA 91125}
\affiliation{$^3$Jet Propulsion Laboratory, California Institute
     of Technology, Pasadena, CA 91109} 

\date{\today}

\begin{abstract}
Baryon-density perturbations of large amplitude may exist
if they are compensated by dark-matter perturbations such that the
total density is unchanged.  Primordial abundances and
galaxy clusters allow these compensated isocurvature perturbations
(CIPs) to have amplitudes as large as $\sim10\%$. CIPs will modulate the
power spectrum of cosmic microwave background (CMB)
fluctuations---those due to the usual adiabatic
perturbations---as a function of position on the sky.  This
leads to correlations between different spherical-harmonic
coefficients of the temperature and/or polarization maps,
and induces polarization $B$-modes.  Here, the magnitude of these effects is calculated and
techniques to measure them are introduced. While a CIP of this amplitude can be probed on large scales with existing data, forthcoming CMB experiments should improve the sensitivity to CIPs by at least an order of magnitude.
\end{abstract}

\pacs{98.80.-k}

\maketitle

We have been conditioned to believe that the $\sim10^{-5}$
variations in the cosmic microwave background (CMB) temperature
\cite{Bennett:1996ce,Komatsu:2010fb} imply that the matter in the
early Universe was distributed
with similarly small variations.  This is certainly true if
primordial perturbations are adiabatic; i.e., if there are
perturbations only to the {\it total} matter content, with the
fractional contributions of baryons, dark matter, photon, and
neutrinos the same everywhere.  It is also true for many 
isocurvature models \cite{Bean:2006qz}, where the total
density is fixed.

It is therefore a surprise that
perturbations in the baryon density can be almost arbitrarily
large---far larger than $10^{-5}$---as long they are
compensated by dark-matter perturbations in such a way that the
total-matter density remains unchanged
\cite{Gordon:2009wx,Holder:2009gd}.  These
compensated isocurvature  perturbations (CIPs) induce no
gravitational fields, as the total matter density in this mode is
spatially homogeneous.  Baryon-pressure gradients
induce motions at the baryon sound speed which, at the time of
primary CMB decoupling ($z\sim 1091$, \textit{decoupling} hereafter), is $(v/c)\sim (T/m_p)^{1/2}\sim
(\mathrm{eV}/\mathrm{GeV})^{1/2}\sim 10^{-4.5}$.
These motions affect the photon temperature
only on distances $\lesssim10^{-4.5}$ times the horizon at
decoupling, that is, CMB multipole moments with $l\gtrsim 10^6$
\cite{Gordon:2009wx}, far larger than those ($l \lesssim 10^4$)
probed by CMB experiments. Thus, while the CMB power spectrum
currently constrains the {\it mean} baryon--to--dark-matter
ratio precisely, it tells us nothing about spatial
variations in this ratio.

Big-bang nucleosynthesis (BBN) and galaxy-cluster baryon fractions
constrain the CIP amplitude to be less than $\sim10\%$
\cite{Holder:2009gd}.  Consequences of CIPs for galaxy surveys
are small \cite{Gordon:2009wx}.  Measurements of 21-cm radiation
from the dark ages would be sensitive to CIPs
\cite{Barkana:2005xu,Lewis:2007kz,Gordon:2009wx}, but such
measurements are still off in the future.
%somewhat in the future. seemed to strange that something IS in the future...tense issue?

Here we show that the primordial relative distribution of
baryons and dark matter can be determined with the CMB.  Our
principle motivation is curiosity---is the common assumption
that baryons trace dark matter in the early Universe justified
empirically?  However, a search for CIPs is also motivated by
the curvaton models \cite{Lyth:2002my} that predict their
existence \cite{Gordon:2009wx,Gordon:2002gv} and perhaps by
recent ideas linking baryon and dark-matter densities
\cite{Kaplan:2009ag}.  Moreover, if the Planck satellite finds
evidence for primordial isocurvature perturbations, the CIP
measurements we describe below will be essential to determine
how that perturbation is distributed between baryons and
dark matter.

CIPs modulate the baryon and dark-matter densities at decoupling, where $\sim 90\%$ of photons last scatter, and at reionization, where $\sim10\%$ of CMB
photons last scatter.  There will thus be a modulation of the
small-scale temperature and polarization power spectra from one
patch of sky to another.  This modifies the power spectrum obtained by
averaging over the entire sky, induces polarization $B$ modes,
and causes correlations between different spherical-harmonic
coefficients of the temperature and/or polarization maps.  The effects
on the CMB are analogous to those of gravitational
lensing \cite{Lewis:2006fu}. The $B$-mode power spectrum induced
by CIPs through the modulation of the \textit{reionization}
optical depth has already been calculated \cite{Holder:2009gd}.

We show, however, that the CMB effects induced by modulation of
the baryon density at \textit{decoupling} are considerably larger than
those induced at reionization.  Our calculation follows
Ref.~\cite{Sigurdson:2003pd}, where the CMB effects of a
spatially-varying cosmological parameter (there the
fine-structure constant) were considered.  This variation induces a spatially-varying power
spectrum.  We have extended the formalism of Ref.
\cite{Sigurdson:2003pd} to calculate the effect of CIPs on top of the usual adiabatic initial conditions, extending the flat-sky formalism developed there to the full sky, and generalizing the calculation to scales of smaller width than the decoupling surface.
Since the technical details are complicated, we present
them elsewhere \cite{Grin:2011tf} and focus here on the
principal science results.

\begin{figure}[t]
\includegraphics[width=3.20in]{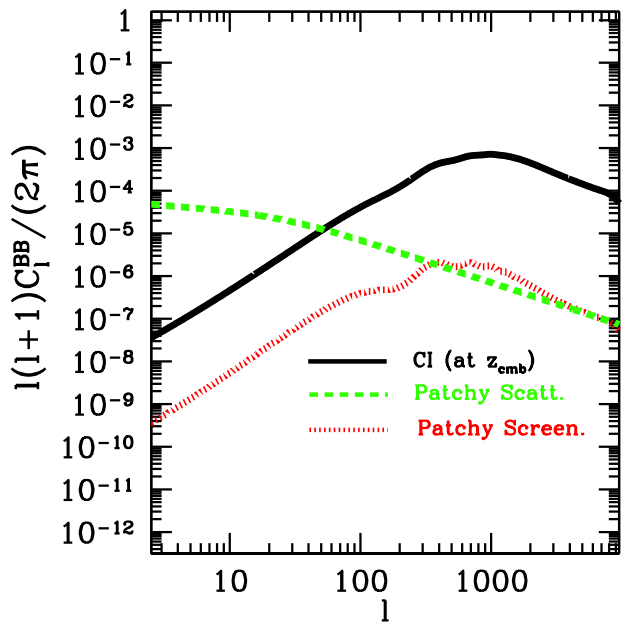}
%or alt_bb_prl.ps
\caption{The $B$-mode power spectrum (in $\mu{\rm K}^{2}$) for the CIP-induced
     contribution from decoupling (solid 
     black curve),  contrasted with contributions from patchy
     scattering (short-dash green curve), and patchy screening (dotted red curve) at reionization. We use a scale-invariant spectrum of CIPs with the
     amplitude $A\simeq0.017$ that saturates the galaxy-cluster
     bound.}
  %    tensor modes (dot-dash cyan curve) with tensor-scalar ratio $r=0.1$, and gravitational lensing of the CMB (purple)
\label{fig:Bmodes}
\end{figure}

%citations
The CIP involves baryon and cold-dark-matter densities
$\rho_b(\bfx)=\bar\rho_b [ 1+ \Delta(\bfx)]$ and $\rho_c(\bfx) =
\bar\rho_c - \rho_b \Delta(\bfx)$, written as
functions of position $\bfx$ in terms of a fractional
baryon-density perturbation $\Delta(\bfx)$.  Note that the
total matter density $\rho_b(\bfx)+\rho_c(\bfx)$ associated with
the CIP does not vary with $\bfx$.  We assume that 
$\Delta(\bfx)$ is a random field with a
scale-invariant power spectrum $P_\Delta(k) = A k^{-3}$, as may
be expected if CIPs arise somehow from inflation, and $A$ is
a dimensionless amplitude.  The rms variation
$\Delta_{\mathrm{cl}}$ in the baryon--to--dark-matter ratio
between galaxy clusters obeys the constraint $\Delta_{\mathrm{cl}}\lsim 0.08$.

\begin{figure*}[t]
\centering
\includegraphics[width=7.00in]{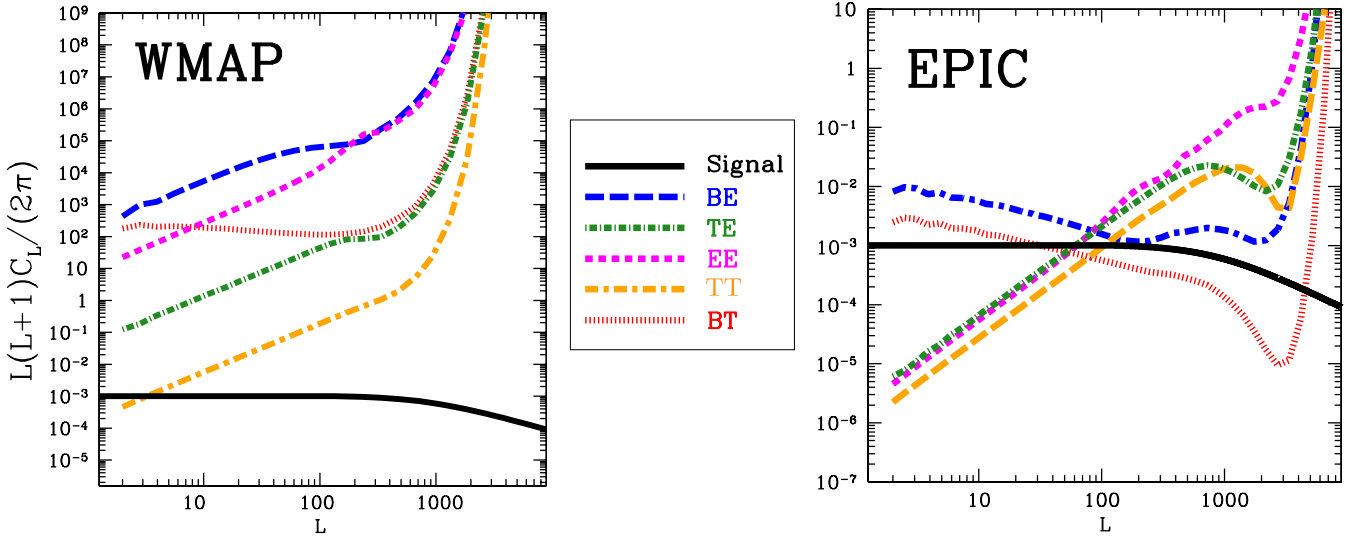}\caption{Shown are the errors in
     $C_L^\Delta$ from the $TT$, $TE$, $EE$, $TB$, and $EB$ estimators for
     the CIP perturbation $\Delta$ at the surface of last
     scatter for (a) WMAP and (b) a CMB polarization satellite, with the
     specifications spelled out in the EPIC mission concept
     study.  Also shown (signal) is the power spectrum
     $C_L^\Delta$ for a scale-invariant spectrum of CIPs with
     the maximum amplitude allowed by galaxy clusters.}
\label{fig:sensitivities}
\end{figure*}

When the three-dimensional field is projected onto a narrow
spherical surface, the resulting angular power spectrum for
$\Delta$ will be $C_L^\Delta\simeq A/(\pi L^2)$ for mulipole moments
$L\lsim (\eta_0-\eta_{\mathrm{ls}})/\sigma_\eta$, where
$\eta_{\mathrm{ls}}$ and $\eta_0$ are the conformal time at last
scatter and today, respectively, and $\sigma_\eta$ is the rms
conformal-time width of the last-scattering surface.  At smaller
angular scales (larger $L$), the variation in $\Delta$
is suppressed by the finite width of the scattering surface.
The angular power spectrum for $\Delta$ can then be approximated
by $C_L^\Delta \simeq A (\eta_0-\eta_{\mathrm{ls}})/(2\sqrt{\pi}
L^3\sigma_\eta)$ for $L\gsim (\eta_0 - \eta_{\mathrm{ls}}) /
\sigma_\eta$ \cite{Holder:2009gd,Grin:2011tf}. The rms variation
$\Delta_{\mathrm{cl}}$ in the baryon--to--dark-matter ratio
on galaxy cluster scales is
\begin{equation}
     \Delta_{\mathrm{cl}}^2 = \frac{1}{2\pi^2} \int k^2 \,
     dk\, \left[ 3 j_1(kR)/(kR) \right]^2 P_\Delta(k),
\label{eqn:clusterrms}
\end{equation}
where $R$ is the mean separation between galaxy
clusters.  The integral has a formal logarithmic divergence at
low $k$ which is cut off, however, by the horizon
$k_{\mathrm{min}}\simeq(10~\mathrm{Gpc})^{-1}$.  Taking
$R\simeq10$~Mpc, we find $\Delta_{\mathrm{cl}}^2\simeq A
\ln(1000)/2\pi^2$.  Thus, $\Delta_{\mathrm{cl}}\lesssim0.08$
implies $A\lesssim 0.017$. A weaker bound ($A\lesssim 0.046$) comes from BBN. \begin{figure*}[t]
\centering
\includegraphics[width=6.90in]{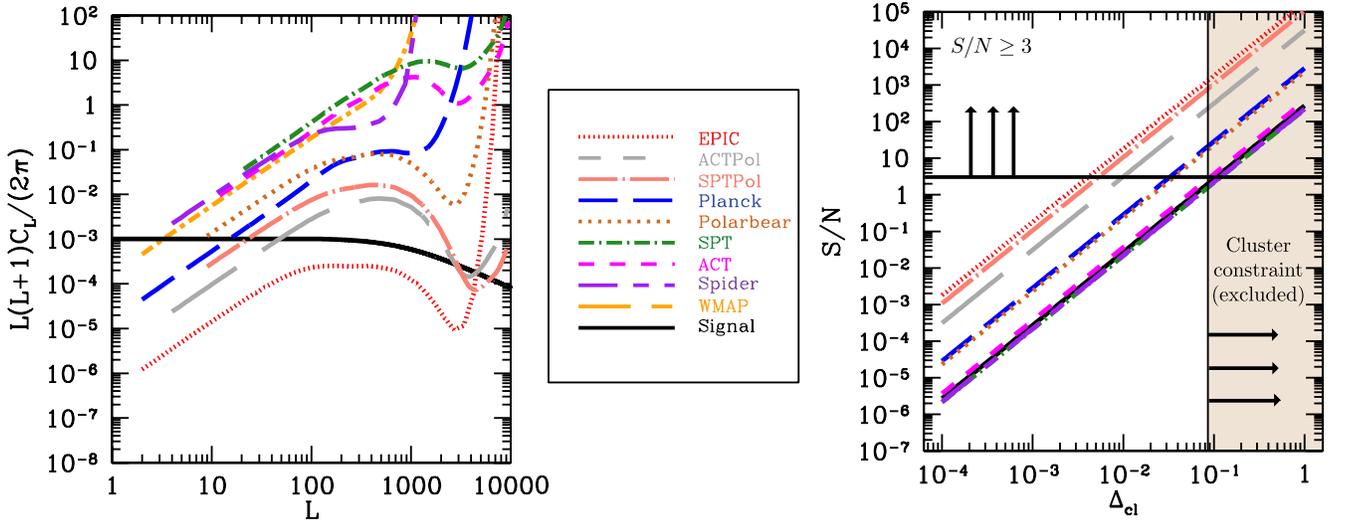} \caption{(a) Shown are the total expected errors in
     $C_L^\Delta$ from the combined $TT$, $TE$, $EE$, $TB$, and $EB$
     estimators for the CIP perturbation $\Delta$ at the surface
     of last scatter for several current and forthcoming CMB
     experiments.  Also shown (signal) is the power spectrum
     $C_L^\Delta$ for a scale-invariant spectrum of CIPs with
     the maximum amplitude allowed by galaxy clusters. (b) The signal-to-noise as a function of the
     rms fluctuation in the galaxy cluster baryon--to--dark-matter ratio along different lines of sight at $z\simeq 1091$.  The vertical line
     shows the upper limit to the rms amplitude from galaxy clusters (excluded region is shaded).}
\label{fig:final}
\end{figure*}

Now consider the CMB fluctuations produced at decoupling. The CIP-induced variation of the baryon and dark-matter densities across the sky modulates the small-scale power spectrum, and this modulation induces off-diagonal correlations in the CMB \cite{Sigurdson:2003pd,Grin:2011tf}.  

Moreover, $B$ modes are induced in the CMB polarization
\cite{Sigurdson:2003pd}.  The induced spherical-harmonic
coefficients are
\begin{eqnarray}
     a_{lm}^{B} = -i \sum_{L+l+l'~\mathrm{odd}}
     \xi^{LM}_{lml'm'} \wigner{l}{2}{L}{0}{l'}{-2} \Delta_{LM}
     \frac{d a_{l'm'}^{E}}{d\Delta}
\label{bb_pspec_new},
\end{eqnarray}
where $\Delta_{LM}$ are the spherical-harmonic coefficients for
$\Delta(\hat n)$; $da^{E}_{l'm'}/d\Delta$ is the derivative of
the usual $E$-mode spherical-harmonic coefficient with respect to
$\Delta$ (computed using the \textsc{camb} code \cite{Grin:2011tf}) and
\begin{eqnarray}
 \xi^{LM}_{lml'm'} = \wigner{l}{0}{L}{0}{l'}{0}^{-1}\int d\hat n Y_{lm}^*(\hat n) Y_{LM}(\hat n)Y_{l'm'}(\hat n).
\label{eqn:xi}
\end{eqnarray} 
This induced $B$ mode arises by modulating the
first-order adiabatic perturbation to first order in the CIP.  This is because the sound speed, photon diffusion length, and visibility function, assumed spatially homogeneous in the standard treatment, all depend on the local baryon density. In contrast, when the CMB is gravitationally lensed,  $d a_{l'm'}^{E}/d\Delta$ is replaced by a function encoding a deflection.

Figure ~\ref{fig:Bmodes} shows the results of our
calculations for the $B$-mode power spectrum $C_l^{\rmB\rmB}$
induced by a scale-invariant spectrum of CIPs with the largest
amplitude ($A\simeq0.017$) consistent with galaxy-cluster
baryon--to--dark-matter ratios.  CIPs modulate the reionization optical depth, and as
noted in Ref.~\cite{Holder:2009gd}, this also generates $B$ modes,
through patchy screening \cite{Dvorkin:2008tf} and
scattering \cite{Hu:1999vq} of primordial CMB fluctuations.
We plot these reionization
contributions in Fig.~\ref{fig:Bmodes} for the same CIP
amplitude.  We see that the $B$-mode power spectrum induced at decoupling
is larger (by up to $3$ orders of magnitude) than that induced at reionization, for $l\gtrsim50$.  The
decoupling-induced $B$ modes are larger because (a) they involve
$\sim90\%$ of the photons, rather than $\sim10\%$, and (b)
the finite width of the reionization re-scattering surface
smooths the angular $\Delta$ fluctuations to larger angular
scales (lower $L$) than it does for decoupling. Reconstruction
of $\Delta_{LM}$ depends primarily on higher-$l$ modes, and so
the baryon-density modulation at decoupling is more important in
probing CIPs than that at reionization. 

%We also compare with the power spectrum of primordial B-modes (for scalar-tensor ratio $r=0.1$) and gravitational lensing induced B-modes. For $l\gtrsim600$, the CIP-induced signal exceeds the tensor-induced signal. Although the CIP-induced signal is smaller than the lensing-induced signal, the detailed form of the off-diagonal correlations induced allows the CIP contribution to be separated from the lensing contribution, as described below. 

We now turn to the reconstruction of
CIPs from CMB maps.  In the absence of CIPs, the multipole
moments $a_{lm}^{X}$ obey the relation $\VEV{a_{lm}^{ X} a_{l'm'}^{X^{\prime}}} = C_l^{X  X^{\prime}} \delta_{ll'}
\delta_{mm'}$ ($X\in\{{T,E,B}\}$); i.e., spherical-harmonic coefficients with $(lm) \neq (l'm')$ are statistically independent. However, if there is spatial modulation of the power spectrum, then there will be off-diagonal ($l\neq
l^{\prime}$, $m\neq m^{\prime}$) correlations,
\begin{align}
     \langle a_{lm}^X a_{l'm'}^{X^{\prime}}\rangle = &~C_l^{ X X^{\prime}}\delta_{ll'}
    \delta_{mm'} \nonumber\\+& \sum_{LM} D_{l l^{\prime}}^{LM,{ X X^{\prime}}}
     \xi^{LM}_{lml'm'},
\label{eqn:offdiagonal}
\end{align}
where $D_{l l^{\prime}}^{LM,{ X X^{\prime}}}=\Delta_{LM}
S_{ll'}^{L,XX^{\prime}}$ are bipolar spherical
harmonics \cite{Hajian:2003qq}, and $S_{ll'}^{L,XX'}$
are coupling coefficients that are
calculated in Ref.~\cite{Grin:2011tf}. 

Construction of minimum-variance estimators for the $\Delta_{LM}$
and their associated errors is
straightforward \cite{Pullen:2007tu}. 
For example, for the $EB$ correlation, the estimator is
\begin{eqnarray}
   \widehat{\Delta}_{LM}&=&\sigma_{\Delta_{L}}^{2}\sum_{l^{\prime}\geq
   l} \frac{S_{l l^{\prime}}^{L,{ EB}}
   \widehat{D}_{l l^{\prime}}^{LM,{EB}}}{C_{l}^{
BB,{\rm map}}C_{l^{\prime}}^{EE,{\rm map}}}+ \left\{{E
\leftrightarrow B} \right\} ,\end{eqnarray} and it has a variance, \begin{equation}
\sigma_{\Delta_{L}}^{-2}=\sum_{l^{\prime}\geq l} \frac{\left(2l+1\right)\left(2l^{\prime}+1\right)\left(S_{l l^{\prime}}^{L,{ EB}}\right)^{2}}{4\pi C_{l}^{BB,{\rm map}}C_{l^{\prime}}^{EE,{\rm map}}}+\left\{ { E \leftrightarrow B} \right\}, 
%\nonumber
 %\\ G_{l l^{\prime}}&\equiv&\frac{\left(2l+1\right)\left(2l^{\prime}+1\right)}{4\pi}.
\label{singleatatime}
\end{equation}
where $C_{l}^{ X X'}$ are power spectra including noise, and $\widehat{D}_{l l'}^{LM,{EB}}$ is the minimum-variance estimator for $D_{l l^{\prime}}^{LM,{
EB}}$ \cite{Pullen:2007tu}.  From these one can estimate $C_L^\Delta$. 

Figure \ref{fig:sensitivities} shows the predicted errors in the CIP power spectrum
reconstruction from the $TT$, $EE$, $TE$, $TB$, and $EB$
estimators for the Wilkinson Microwave Anisotropy Probe (WMAP) and the proposed Experimental Probe of Inflationary Cosmology (EPIC).  These are $\delta C_{L}^{ X X'}\equiv \left[(2L+1)\right]^{-1/2} (\sigma_{\Delta_{L}}^{ X X'})^{2}/f_{\rm sky}$, where the $(2L+1)^{-1/2}$ factor results from the multiple modes available at each $L$.

Instrumental parameters for WMAP are
a beamwidth of $21'$ (full-width half-max), noise-equivalent temperature (NET) of $1200~\mu{\rm K}\sqrt{\rm s}$, fraction of sky analyzed $f_{\rm sky}=0.65$, and observation time $t_{\rm obs}=7~{\rm years}$.  For EPIC (150 Ghz channel) we assume a beamwidth of $5$ arcmin, NET of $2.0~\mu{\rm K}\sqrt{\rm s}$, and observation time $t_{\rm obs}=4~{\rm years}$, also with $f_{\rm sky}=0.65$. 

For WMAP, the best sensitivity comes from $TT$.  For EPIC, the
sensitivity at $L \gtrsim 100$ comes primarily from the $TB$
estimator.  We have checked that the best sensitivity for Planck
comes from $TT$, while some ground-based experiments (e.g.,
SPTPol) benefit from polarization. 

The left panel of Fig.~\ref{fig:final} shows the errors in the CIP power spectrum
reconstruction obtained by combining the $TT$,
$TE$, $EE$, $TB$, and $EB$ estimators for a variety of CMB experiments. The signal to noise ratio is given by
$S/N=\{(f_{\rm sky}/2)\sum_{L>f^{-1/2}_{\rm sky}}\left(2L+1\right)[(C_{L}^{\Delta}/\sigma_{\Delta_L}^{2})]^{2}   \}^{1/2}$. The right panel shows the $S/N$ for detection of a
scale-invariant spectrum of CIPs as a function of $\Delta_{\rm cl}$.  With WMAP, a CIP saturating the cluster bound is marginally accessible on the largest scales. Planck should be able to probe rms CIP amplitudes of $3\times 10^{-2}$ and higher. Significant improvements in sensitivity should be obtained with upcoming experiments like Polarbear, SPTPol, and ACTPol. We see that $S/N$ values $\gtrsim 3$ may be possible with EPIC for an rms CIP amplitude of $4\times 10^{-3}$, a factor of $\sim 20$ lower than the current limit.

The tools for these measurements should be generalizations of those used for weak lensing of the CMB \cite{Smith:2007rg}, which also produces off-diagonal correlations. CIPs should be distinguishable from lensing, since these physical effects are distinct, as evidenced by differing forms for  the coupling coefficients $S_{l l^{\prime}}^{L, X X^{\prime}}$. In Ref. \cite{Su:2011ff}, it is shown that for the analogous case of patchy reionization, optical depth fluctuations may be separated from lensing without significant loss in $S/N$, and we expect that this is also true for CIPs. We leave for future work the development of tools to distinguish CIPs from weak lensing and contaminants like Galactic foregrounds.

CIPs are an intriguing possibility and a prediction of some
inflationary models.  With the measurements we have described
here, we may soon know empirically how closely dark matter and
baryons trace each other in the early Universe. 

\bigskip
We thank G. Holder, T.~L.~Smith, M.~LoVerde, C.~Chiang, K.~M.~Smith, M.~Zaldarriaga, and D.~N.~Spergel for stimulating conversations. We thank B.~Jones and A. Fraisse for providing Spider parameters. DG was supported by NSF
AST-0807044 and MK by DoE DE-FG03-92-ER40701 and NASA
NNX10AD04G. Part of this research was carried out at the Jet
Propulsion Laboratory, Caltech, under a contract with NASA.

\end{document}